\documentclass[onecolumn,11pt,a4paper]{revtex4}

\newcommand{\eq}[1]{\begin{equation}#1\end{equation}}
\newcommand{\dd}{\mathrm{d}}
\newcommand{\ee}{\mathrm{e}}
\usepackage{amsmath}
\usepackage{graphics}
\usepackage{graphicx}

\begin{document}

\title{Entanglement in a periodic quench}
\author{Viktor Eisler\footnote{Corresponding author \quad E-mail: {\sf eisler@physik.fu-berlin.de}} and Ingo Peschel}
\address{Fachbereich Physik, Freie Universit\"at Berlin, Arnimallee
14, 14195 Berlin, Germany}

%\keywords{Entanglement entropy, fermionic chains, quantum dynamics.}
%\subjclass[pacs]{02.30.Ik, 03.67.Bg, 05.70.Ln}

\begin{abstract}
We consider a chain of free electrons with periodically switched dimerization
and study the entanglement entropy of a segment with the remainder of the
system. We show that it evolves in a stepwise manner towards a value
proportional to the length of the segment and displays in general slow
oscillations. For particular quench periods and full dimerization an explicit
solution is given. Relations to equilibrium lattice models are pointed out.
\end{abstract}

\maketitle

\section{Introduction}

Entanglement is a basic feature of non-trivial quantum states and has been
studied extensively for the ground states of standard many-body systems
\cite{Amicoetal07}. However, the question how it evolves in time-dependent
situations is equally interesting. The simplest case is a quench where the 
Hamiltonian is changed instantaneously from $H_0$ to $H_1$ and one follows the 
subsequent evolution of the initial state, usually taken to be the ground state 
of $H_0$. For global quenches, where a parameter is changed everywhere in the 
same way, it was found that in homogeneous chains the entanglement entropy $S$
between a segment of length $L$ and the remainder first increases linearly with
time and then approaches an asymptotic value proportional to $L$ \cite{CC05,Chiara06,Schuch08,Laeuchli08}.
Thus the time evolution leads to a much larger entanglement than found in 
equilibrium where even for critical chains $S$ is only proportional to $\ln L$.
This result can be interpreted in a simple picture where pairs of quasiparticles
establish the entanglement between the two parts of the system \cite{CC05}.
Alternatively, the time-dependent state may be viewed as the ground state of an
evolving Hamiltonian which is strongly non-local and gives long-ranged correlations.
\par
In this paper we will study a more general situation and consider not a single
quench, but a periodic sequence of changes $H_0 \leftrightarrow H_1$ and its effect
on the entanglement entropy. Such periodic changes have
been considered in some papers on Ising chains kicked by a transverse 
\cite{Prosen00,LS05,BMM05} or a tilted field \cite{Prosen02,LS05,Prosen07}. 
We will investigate a system of
free electrons hopping on a chain with alternating bonds. The corresponding
dimerization parameter is then switched between the values $\pm \delta$, which
simply interchanges weak and strong bonds. Although there is a mapping to the
transverse Ising chain, the hopping model has a more direct physical significance
since it could be realized with optical lattices. There have already been
experiments with dilute gases where periodic potentials were applied
in the form of kicks \cite{Saunders07}. Other experiments 
\cite{Stoeferle04,Schori04} have motivated studies of chains
where the parameters in the Hamiltonian were assumed to
vary harmonically. In this case, however, the focus was either
on the energy absorption \cite{Iucci06,Kollath06,Kollath06b}
or on work fluctuations \cite{Dorosz07}.
\par
The periodic quench has some important general features. In the evolution,
one can distinguish what happens after full quench periods and what happens at
the times in between. This situation is analogous to the case of a particle moving 
in a periodic potential and has been discussed extensively for simple kicked
quantum systems \cite{Haake00}. For a full quench period, the time evolution
operator is the product of two exponentials and analogous to the transfer 
matrix in a classical two-dimensional system with a layered structure.
One can write it as a single exponential with a certain effective Hamiltonian.
The formulae are very similar to those one finds for the Ising model on a square 
lattice. They also show that the effective Hamiltonian is critical if, as we assume, 
$H_0$ and $H_1$ act for the same length of time. Compared to a simple critical 
quench, however, one has the additional evolution during the intermediate times.
The unitarity of the discrete evolution from period to period, on the other hand,
leads to effects not present in models with Euclidian time.
\par
We describe the model and the method of computing $S$ via correlation functions
in Section 2. Then we treat in Section 3 the fully dimerized case $\delta=1$
which is particularly simple. The time evolution can then be represented 
graphically and for special values of the quench period the particles even 
move with strictly uniform velocity. For $S(t)$ one finds a step structure and an 
asymptotic value proportional to $L$. We discuss the dependence on the quench period 
and show the behaviour for arbitrary times. Section 4 deals with arbitrary 
dimerization. Here the quasiparticles in the effective Hamiltonian can have a more 
complicated dispersion in which case additional slow oscillations appear in the
entanglement entropy. In Section 5 we sum up our findings and in the Appendix 
we give the formulae for calculating the correlation matrix in the general case. 

\section{Model and method}
\label{sec:model}

\begin{figure}[thb]
\center
\includegraphics[scale=.6]{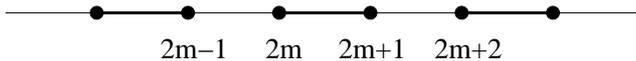}
\caption{Geometry of the dimerized hopping model.}
\label{fig:dimchain}
\end{figure}

We will study a half-filled ring of free electrons with $2N$ sites and
alternating hopping as shown in Fig. \ref{fig:dimchain}.
%
%%%%%%%%%%%%%%%%%%%%%%%%%%%%%%%%%%%%%%%
%%%%%%%%%%%%%%%%%%%%%%%%%%%%%%%%%%%%%%%
\par \noindent
The Hamiltonian is
\eq{
%H = -J \sum_{m=1}^N \left[ \frac{1+\delta}{2}
%(c_{2m}^\dag c_{2m+1}+c_{2m+1}^\dag c_{2m})+ \frac{1-\delta}{2}
%(c_{2m+1}^\dag c_{2m+2}+c_{2m+2}^\dag c_{2m+1}) \right]}
H = -J \sum_{m=1}^N \left( \frac{1+\delta}{2} \, c_{2m}^\dag c_{2m+1}
+ \frac{1-\delta}{2} \, c_{2m+1}^\dag c_{2m+2} + \mathrm{h.c.} \right)
\label{eq:h}}
and, up to boundary effects, also describes a dimerized $XX$ spin model.
In the following we set $J=1$. The diagonal form of $H$ is obtained by working
with cells of two sites, defining $c_{2m}=a_m$, $c_{2m+1}=b_m$, and
performing a Fourier transformation
\eq{
\left(\begin{array}{c} a_m \\ b_m \end{array}\right) =
\frac{1}{\sqrt{N}} \sum_q \ee^{iqm}
\left(\begin{array}{c} a_q \\ b_q \end{array}\right) \; ,
\quad q=\frac{2\pi}{N}k \; .
\label{eq:ab_ft}}
Then
\eq{
H = -\sum_q \omega_q (\alpha_q^\dag \alpha_q - \beta_q^\dag \beta_q)
\label{eq:hdiag}}
where
\eq{
\omega_q = \sqrt{\cos^2 \frac{q}{2}+\delta^2 \sin^2 \frac{q}{2}}
\label{eq:omegaq}}
and the Fermi operators are related via
\eq{
a_q=\frac{1}{\sqrt{2}} \ee^{-i\varphi_q /2}(\alpha_q + \beta_q)
\quad,\quad
b_q=\frac{1}{\sqrt{2}} \ee^{i\varphi_q /2}(\alpha_q - \beta_q)
\label{eq:alpha_beta}}
with
\eq{
\ee^{i\varphi_q} = \ee^{iq/2} \,
\frac{\cos \frac{q}{2}- i\delta \sin \frac{q}{2}}{\omega_q} \; .
\label{eq:phiq}}
\par
In the ground state, all $\alpha$-levels are occupied and all
$\beta$-levels are empty. The single-particle spectrum has a gap
at $q=\pm \pi$ for all $\delta \ne 0$. For vanishing dimerization
$\delta$, (\ref{eq:omegaq}) gives $\omega_q =\cos (q/2)$ which is the
usual result for homogeneous hopping, but folded into the smaller
Brillouin zone. For $\delta = 1$ the dispersion becomes flat,
$\omega_q = 1$. A change $\delta \to -\delta$ does not affect
$\omega_q$ which simplifies the situation in the quench.
\par
At time $t=0$, the system is taken to be in the ground state
$|\Phi_0 \rangle$ of $H_0 = H(\delta)$. The periodic quench
consists of switches between $H_0$ and $H_1=H(-\delta)$ at times
$n \tau$ where $n=0,1,2,\dots$. This interchanges weak and strong
bonds and  forces the system to adapt. Formally, the wave function
then evolves according to
\eq{
|\Phi(t) \rangle = U(t) |\Phi_0 \rangle
\label{eq:phit}}
with a unitary operator $U(t)$ which for $t = n \cdot 2\tau$, i.e. after
$n$ full periods, is given by
\eq{
U(2n\tau)=U^n
\label{eq:u_2ntau}}
where
\eq{
U = U_0 \, U_1 = \ee^{-i H_0 \tau} \ee^{-i H_1 \tau}
\label{eq:u_2tau}}
describes the evolution over one period.
\par
We are interested in the entanglement of a segment of $L$ sites with
the remainder of the system. The corresponding entanglement entropy $S$
follows from the reduced density matrix $\rho$ of the segment which has
the form \cite{Peschel03}
\eq{
\rho = \frac{1}{Z}\; e^{-\cal{H}} \; , \quad
{\cal{H}} = \sum_{k=1}^{L} \varepsilon_k(t) f_k^{\dagger} f_k \; .
\label{eq:rho}}
Here $Z$ is a normalization constant ensuring $\mathrm{Tr}\,\rho = 1$
and the fermionic operators $f_k$ follow from the $c_n$ by an orthogonal
transformation. Then $S= -\mathrm{Tr}\,(\rho \ln \rho)$ is determined by
the single-particle eigenvalues $\varepsilon_k(t)$ according to
\eq{
S(t)=-\sum_k \zeta_k(t) \ln \zeta_k(t) -
\sum_k (1-\zeta_k(t)) \ln (1-\zeta_k(t))
\label{eq:entropy}}
where $\zeta_k(t)=1/(\exp(\varepsilon_k(t))+1)$. The $\zeta_k(t)$ are the
eigenvalues of the correlation matrix
\eq{
C_{lm}(t)=
\langle \Phi(t) | c_l^\dag c_m | \Phi(t) \rangle =
\langle \Phi_0 | c_l^\dag (t) c_m (t) | \Phi_0 \rangle
\label{eq:correlt}}
restricted to the sites of the subsystem. To obtain $\mathbf{C}(t)$
one relates it to $\mathbf{C}(0)$ by expressing the Heisenberg operators
$c_m (t) = U^\dag(t) c_m U(t)$ as linear combinations of the $c_j$.
For this, one has to diagonalize the operator $U$ in (\ref{eq:u_2tau})
which is the product of two exponentials. This is a well-known problem
which first appeared (without the factors of $i$) in the solution of
the two-dimensional Ising model via transfer matrices \cite{SML64}.
A special case is treated in Section \ref{sec:fulldim} while the general
expressions for the elements of $\mathbf{C}(t)$ are given in the Appendix.
Evaluating them numerically and diagonalizing the matrix one then finds
the entropy.
\par
One should note that there is a close relation between $XY$
and transverse Ising (TI) chains \cite{Turban84,PeScho84} which was 
recently extended to their entanglement properties \cite{IJ08,EKPP08}. 
This means that the Hamiltonian $H$ in
(\ref{eq:h}) can be obtained by putting two TI models on alternating
sites of a chain, performing a dual transformation and choosing the
coupling constants $\lambda_\alpha$ and the fields $h_\alpha$ as
$h_1=\lambda_2=(1+\delta)/2$, $h_2=\lambda_1=(1-\delta)/2$. Thus one
TI model is in the ordered and the other in the disordered phase.
Changing the sign of $\delta$ then switches each model to the opposite
phase. In the special case $\delta=\pm 1$ one has either only a field
or only a nearest neighbour coupling. This corresponds to an Ising chain 
kicked periodically by a transverse field \cite{Prosen00,LS05,BMM05}, 
because during the $\delta$-pulses describing the kicks one can
neglect the coupling terms. One could work in this TI formulation, but the 
dimerized chain provides a nicer physical picture as will be seen in the 
following section where we actually treat the case $\delta=1$ before 
dealing with the general situation.

\section{Fully dimerized case}
\label{sec:fulldim}

In this section, we treat a periodic quench between $\delta=\pm 1$
which is particularly transparent and instructive.
This situation has also been considered in \cite{Isart07} for bosons.

\subsection{Time evolution operator}

\begin{figure}[htb]
\center
\includegraphics[scale=.5]{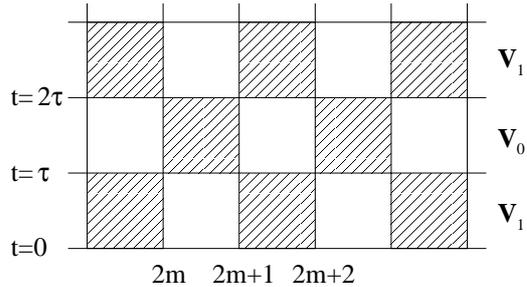}
\caption{Time-evolution pattern for full dimerization. The shaded squares
indicate the coupling of the sites and of the corresponding Fermi operators 
in each half-period.}
\label{fig:chess1}
\end{figure}

For $\delta=1$, the system consists of independent pairs of sites.
For each pair, the Hamiltonian has the form
\eq{
h=-(a^\dag b + b^\dag a)
\label{eq:hfull}}
and is diagonalized by setting $a=(\alpha+\beta)/\sqrt{2}$,
$b=(\alpha-\beta)/\sqrt{2}$. This corresponds to $\omega_q=1$,
$\varphi_q=0$ in (\ref{eq:omegaq})--(\ref{eq:phiq}) and gives
\eq{
h=-(\alpha^\dag \alpha - \beta^\dag \beta) \; .
\label{eq:hfulldiag}}
The time dependence of $a$ and $b$ is then
\eq{
\left(\begin{array}{c} a(t) \\ b(t) \end{array}\right) =
\left(\begin{array}{cc}
\cos t & i \sin t \\
i \sin t & \cos t
\end{array}\right)
\left(\begin{array}{c} a \\ b \end{array}\right)=
v(t) \left(\begin{array}{c} a \\ b \end{array}\right)
\label{eq:abtfull}}
and the full evolution matrix has block form with $N$ such $v$'s
along the diagonal. The same holds for the other half-period
but then the pairs and thus the $v$'s are shifted by one site.
The complete time evolution can therefore be represented as in
Fig. \ref{fig:chess1}, where each shaded square corresponds to a
matrix $v(\tau)$.
%
%%%%%%%%%%%%%%%%%%%%%%%%%%%%%%%%%%%%%%%
%%%%%%%%%%%%%%%%%%%%%%%%%%%%%%%%%%%%%%%
%
Such a chequerboard structure also appears in the time evolution of lattice-gas models 
with parallel dynamics \cite{KDN90,Sch93,HP97}.
In equilibrium problems it is found for partition functions
of vertex models oriented diagonally \cite{TL71} or of one-dimensional quantum 
systems treated with a Trotter decomposition of $\exp(-\beta H)$ \cite{Assaad08}. 
\par
Using translational invariance, the ($2N \times 2N$) matrix 
$\mathbf{V}=\mathbf{V}_0 \mathbf{V}_1$ for one period is found
to have the eigenvalues $\exp(\pm i \gamma_q)$ where
\eq{
\cos \gamma_q = \cos^2 \tau - \sin^2 \tau \cos q \; .
\label{eq:gammaqfull}}
Therefore, writing the operator $U$ in (\ref{eq:u_2tau}) as a single
exponent
\eq{
U= \ee^{-i \bar{H} 2\tau}
\label{eq:u_2tau_full}}
the diagonal form of the effective Hamiltonian is
\eq{
\bar{H} = \sum_q \nu_q (\xi_q^\dag \xi_q - \eta_q^\dag \eta_q)
\label{eq:heff_full}}
where $\nu_q =\gamma_q/2\tau$ and the Fermi operators $\xi_q$, $\eta_q$
follow from the eigenvectors of $\mathbf{V}$.
\par
\begin{figure}[htb]
\center
\includegraphics[scale=.6]{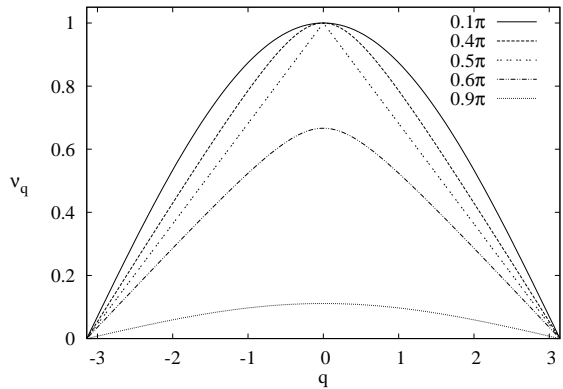}
%\vspace{-0.3cm}
\caption{Dispersion of the single-particle excitations in $\bar{H}$ for
five values of the half-period $\tau$.}
\label{fig:nuq_fulld}
\end{figure}

The single-particle energies of $\bar{H}$ are shown in Fig. \ref{fig:nuq_fulld}
for several values of $\tau$.
%
%%%%%%%%%%%%%%%%%%%%%%%%%%%%%%%%%%%%%%%
%%%%%%%%%%%%%%%%%%%%%%%%%%%%%%%%%%%%%%%
%
The main feature is that the spectrum is always gapless at $q=\pm\pi$.
Thus the evolution over one period is given by a critical Hamiltonian.
This is a consequence of the symmetry of the two half-periods.
If they have different lengths $\tau_0$, $\tau_1$, Eq. (\ref{eq:gammaqfull})
is changed to
\eq{
\cos \gamma_q = \cos \tau_0 \cos \tau_1 - \sin \tau_0 \sin \tau_1 \cos q
\label{eq:gammaqfull_nonsymm}}
and $\gamma_\pi = |\tau_0-\tau_1|$. One should note that the relation
(\ref{eq:gammaqfull_nonsymm}) is the trigonometric analogue of the
hyperbolic formula
\eq{
\cosh \gamma_q = \cosh 2K^*_1 \cosh 2K_2 -
\sinh 2K^*_1 \sinh 2K_2 \cos q
\label{eq:gammaq_hyp}}
found for the row transfer matrix of the two-dimensional Ising model on a square
lattice with couplings $K_1$ and $K_2$ \cite{SML64}. The quantity $K^*_1$ in 
(\ref{eq:gammaq_hyp}) is the
dual coupling of $K_1$, $\tanh K^*_1= \exp(-2K_1)$. This illustrates the close
connection between the two problems mentioned earlier.

\subsection{Hamiltonian limit}

For rapid switching, $\tau \ll 1$, one finds from (\ref{eq:gammaqfull})
$\gamma_q=2\tau\cos \frac q 2$, i.e. $\nu_q = \cos \frac q 2$.
According to (\ref{eq:omegaq}) these are the single-particle energies
$\omega_q$ of $H$ in the absence of dimerization. This is not surprising
since for $\tau \ll 1$ one can write directly
\eq{
U = \ee^{-i H_0 \tau} \ee^{-i H_1 \tau} \approx \ee^{-i(H_0+H_1)\tau}
\label{eq:hamlimit}}
which gives $\bar{H}=(H_0+H_1)/2=H(\delta=0)$. Therefore the evolution
for times $t = n \cdot 2\tau$ is the same as for a single quench to
a homogeneous hopping model. This situation has already been studied in
\cite{EP07}, Appendix B, where the correlation matrix $\mathbf{C}(t)$ was
calculated and the entanglement entropy was found to approach the
asymptotic value
\eq{
S(\infty)=L \, (2\ln 2 - 1)
\label{eq:ent1q_asympt}}
for large $L$.

\subsection{The special case $\tau = \pi/2$}
\label{sec:pihalf}

\begin{figure}[htb]
\center
\includegraphics[scale=.5]{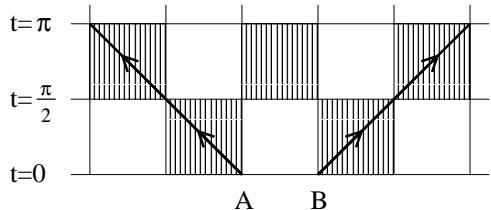}
\caption{Motion of the particles for $\tau= \pi/2$.}
\label{fig:chess2}
\end{figure}

For this value of $\tau$, $\nu_q = 1 \pm q/\pi$ is strictly linear
and describes massless particles with velocity $v=2/\pi$ in units of
the lattice spacing. This can also be seen directly in real space.
According to (\ref{eq:abtfull}), one has
\eq{
a(\frac{\pi}{2})= ib \quad , \quad b(\frac{\pi}{2}) = ia
\label{eq:abpihalf}}
such that particles in a cell interchange positions. This continues in
subsequent half-periods and leads to straight paths as sketched in 
Fig. \ref{fig:chess2}.
%
%%%%%%%%%%%%%%%%%%%%%%%%%%%%%%%%%%%%%%%
%%%%%%%%%%%%%%%%%%%%%%%%%%%%%%%%%%%%%%%
\par
A correlation which exists initially between the first neighbours
A and B is thereby extended to third neighbours in the first step,
to fifth neighbours in the second one and so forth. This permits to
set up the correlation matrix $\mathbf{C}(t)$. If A and B form a cell,
then in the ground state all correlation functions are equal
\eq{
\langle a^\dag a \rangle = \langle a^\dag b \rangle =
\langle b^\dag a \rangle = \langle b^\dag b \rangle = \frac 1 2
\label{eq:correlfull_gs}}
and the same holds then for correlated sites at later times. For the
entanglement entropy only those correlations are relevant which connect
sites inside and outside the subsystem. They lead to an
eigenvalue $\zeta = 1/2$ which gives a contribution of $\ln 2$ to $S$. It is 
not difficult to see that in one step the number of in-out correlations
does not change while in the other one it increases by $4$.
Thus in a full period
\eq{
\Delta S = 4 \ln 2
\label{eq:deltas}}
and the increase continues until $S$ reaches the highest possible 
value $L \ln 2$. This build-up of the entanglement is an exact lattice realization 
of the light-cone picture introduced by Calabrese and Cardy \cite{CC05} in the context 
of single quenches and analyzed further in \cite{Eisert06,Bravyi06}.
\par
\begin{figure}[htb]
\center
\includegraphics[scale=.7]{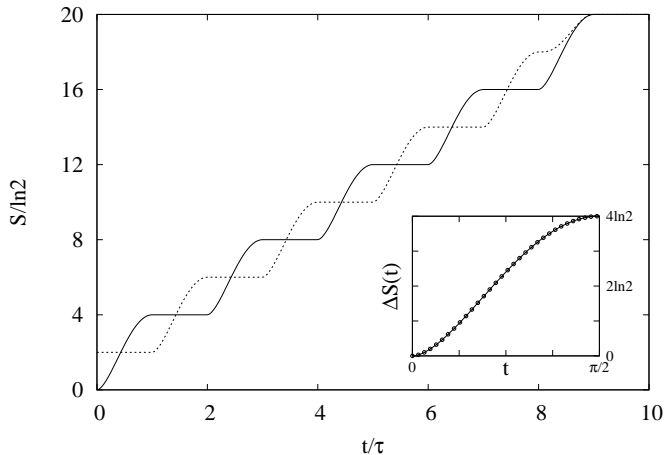}
%\vspace{-0.3cm}
\caption{Entanglement entropy $S(t)$ for $\tau = \pi/2$, $L=20$ and the
two choices of the subsystem, see text. Inset : Enlarged view of the
ascending part $\Delta S(t)$. The line is the analytical result.} 
\label{fig:entpihalf}
\end{figure}

The description becomes complete if one also looks at intermediate times.
By studying small values of $L$ one finds that during the half-period
where $S$ changes, the correlation matrix has non-trivial eigenvalues
$\zeta_1 = \sin^2 (t/2)$ and $\zeta_2 = \cos^2 (t/2)$.
With (\ref{eq:entropy}) these give a contribution
\eq{
\Delta S(t) = -4 \left[ \sin^2 \frac t 2 \ln \sin^2 \frac t 2 +
\cos^2 \frac t 2 \ln \cos^2 \frac t 2 \right]
\label{eq:deltast}}
which describes the increase by $4\ln 2$ between $t=0$ and $t=\pi/2$.
Therefore $S(t)$ has a step-like structure where plateaus are connected
by the function $\Delta S(t)$. The result of a numerical calculation for $L=20$ is 
shown in Fig. \ref{fig:entpihalf}.
Due to the alternating structure of the chain, there are two possible 
choices for the subsystem if $L$ is even. It may contain initially $L/2$
complete cells, in which case it is uncorrelated with the surrounding and $S(0)=0$.
Or it contains only half a cell at each end, which contributes
$\ln2$ to the entanglement giving altogether $S(0)=2\ln2$. This leads to the two curves in the
figure. Note that for the second choice the last plateau is absent and the
final step has only height $2\ln2$. Its form is also slightly modified. The inset shows a 
comparison of $\Delta S(t)$ with the analytical formula (\ref{eq:deltast}).
%
%%%%%%%%%%%%%%%%%%%%%%%%%%%%%%%%%%%%%%%
%%%%%%%%%%%%%%%%%%%%%%%%%%%%%%%%%%%%%%%
\par
These considerations can also be extended immediately to times $\tau$ which are 
multiples of $\pi/2$. Then the particles change their position in the cells several 
times during one half-period. Consequently, the plateaus are longer and $S(t)$ 
follows the function (\ref{eq:deltast}) for a longer interval. Thus 
if $\tau = (2k+1)\pi/2$ the plateaus are connected by an oscillating function.
Such a case is also discussed below for more general times. On the other hand, 
if $\tau$ is an even multiple of $\pi/2$, the oscillations always lead back to 
the initial value of $S$ and there is no systematic ascent.

\subsection{General times $\tau$}
\label{sec:gentau}

\begin{figure}[htb]
\center
\includegraphics[scale=.63]{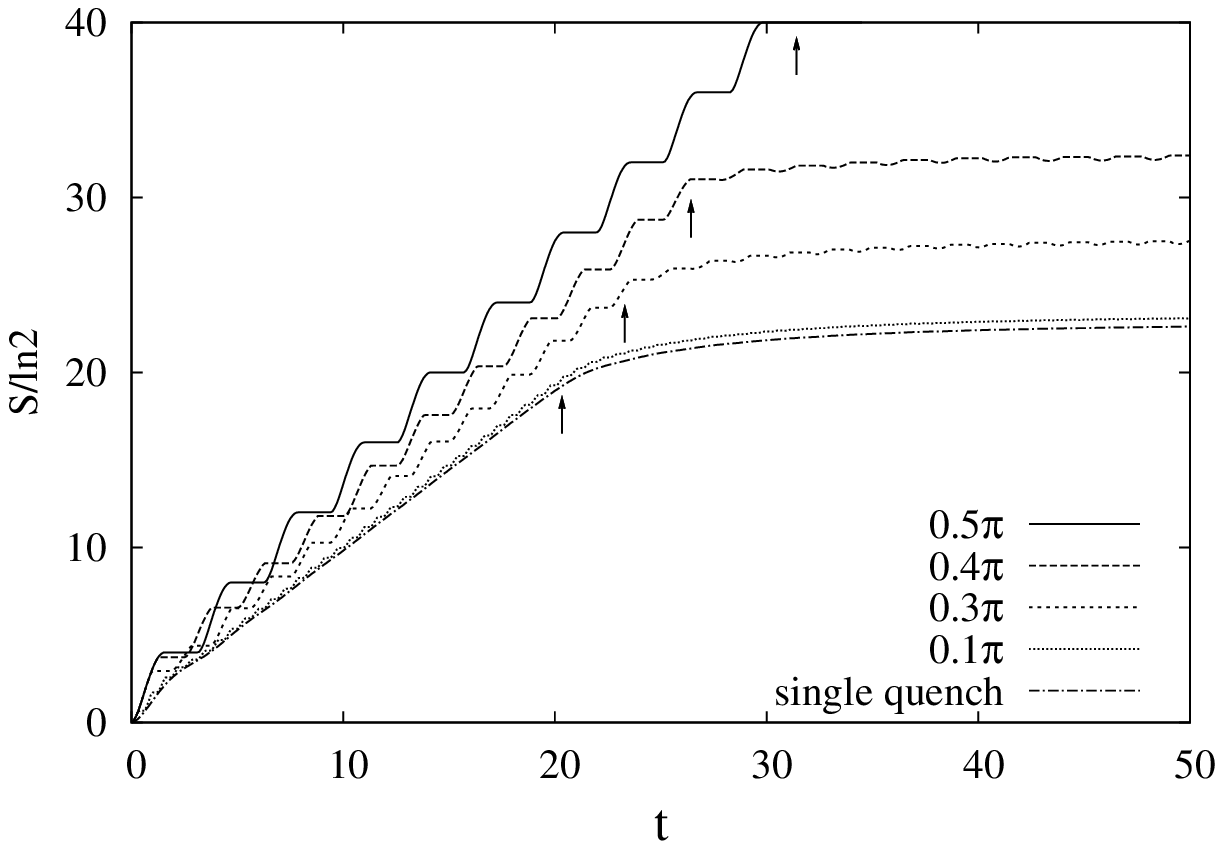}
\includegraphics[scale=.63]{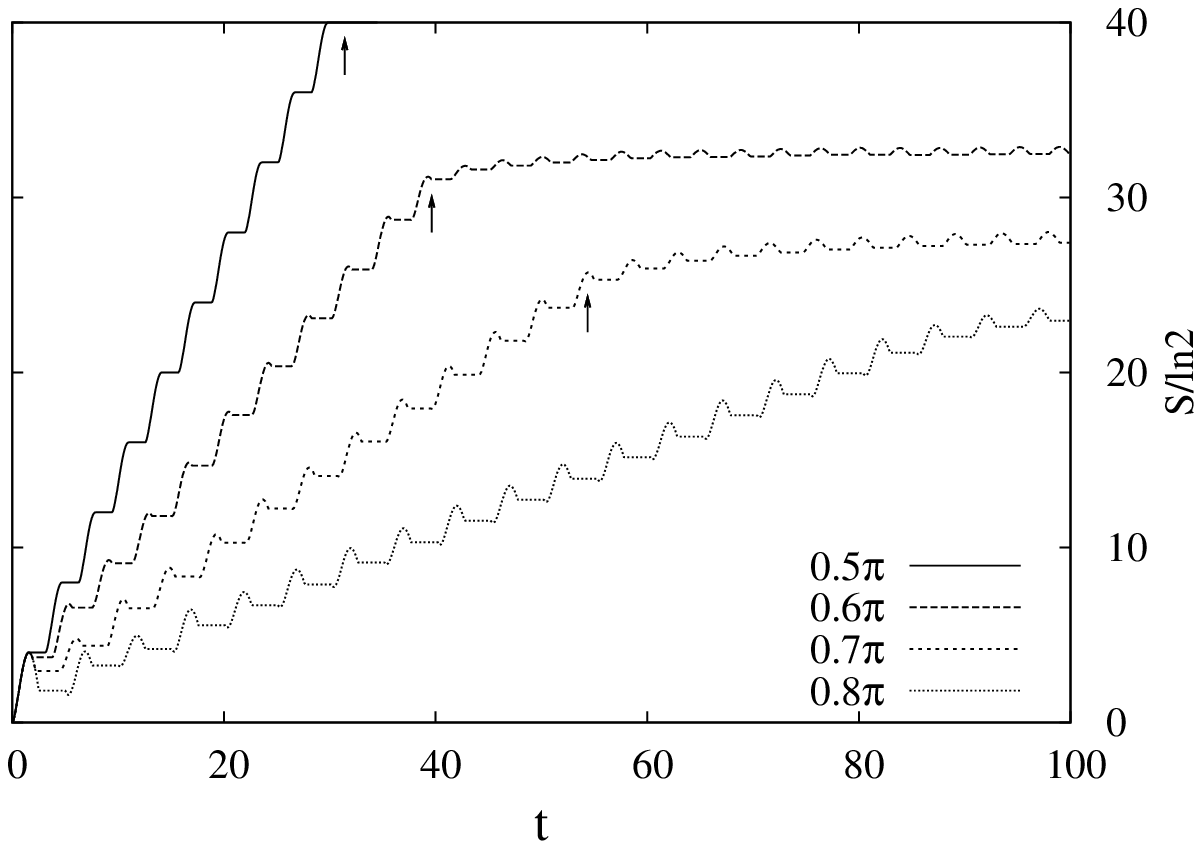}
%\vspace{-0.3cm}
\caption{Entanglement entropy $S(t)$ for $L=40$ and different values 
of $\tau$. Left: Small $\tau$. Right: Larger $\tau$. The arrows indicate
the times $L/2v_{\rm max}$.}  
\label{fig:entfulldtau}
\end{figure}

For general values of the quench period, no analytical treatment is possible
and the entropy has to be obtained numerically as described in Section 2.
Figure \ref{fig:entfulldtau} shows results for small values of $\tau$ on the 
left and for larger ones on the right.
%
%%%%%%%%%%%%%%%%%%%%%%%%%%%%%%%%%%%%%%%
%%%%%%%%%%%%%%%%%%%%%%%%%%%%%%%%%%%%%%%
%
\par

The overall behaviour is an initial, on average linear increase followed by
a sharp bend and a final approach to an asymptotic value. The upper bound
$L \ln2$ is only reached for $\tau= \pi/2$.
The bending points can be related to the
velocity of the fastest one-particle excitations in the effective
Hamiltonian. This is given by
\eq{
v_{\rm max}= 2\left| \frac{\partial \nu_q}{\partial q} \right|_{q=\pi}
= \frac{\sin \tau}{\tau}
\label{eq:vmax}}
where the factor of $2$ is needed for the distance to be measured in sites
instead of cells.
Then the time required for these excitation to travel from the
middle of the block to the surrounding is given by $L/2v_{\rm max}$.
The corresponding values are indicated by arrows in Fig. \ref{fig:entfulldtau}
and locate the cross-over points very well.
In the Hamiltonian limit $\tau \rightarrow 0$, the 
result for the single quench to the homogeneous chain is recovered. 

\par
\begin{figure}[thb]
\center
\includegraphics[scale=.53]{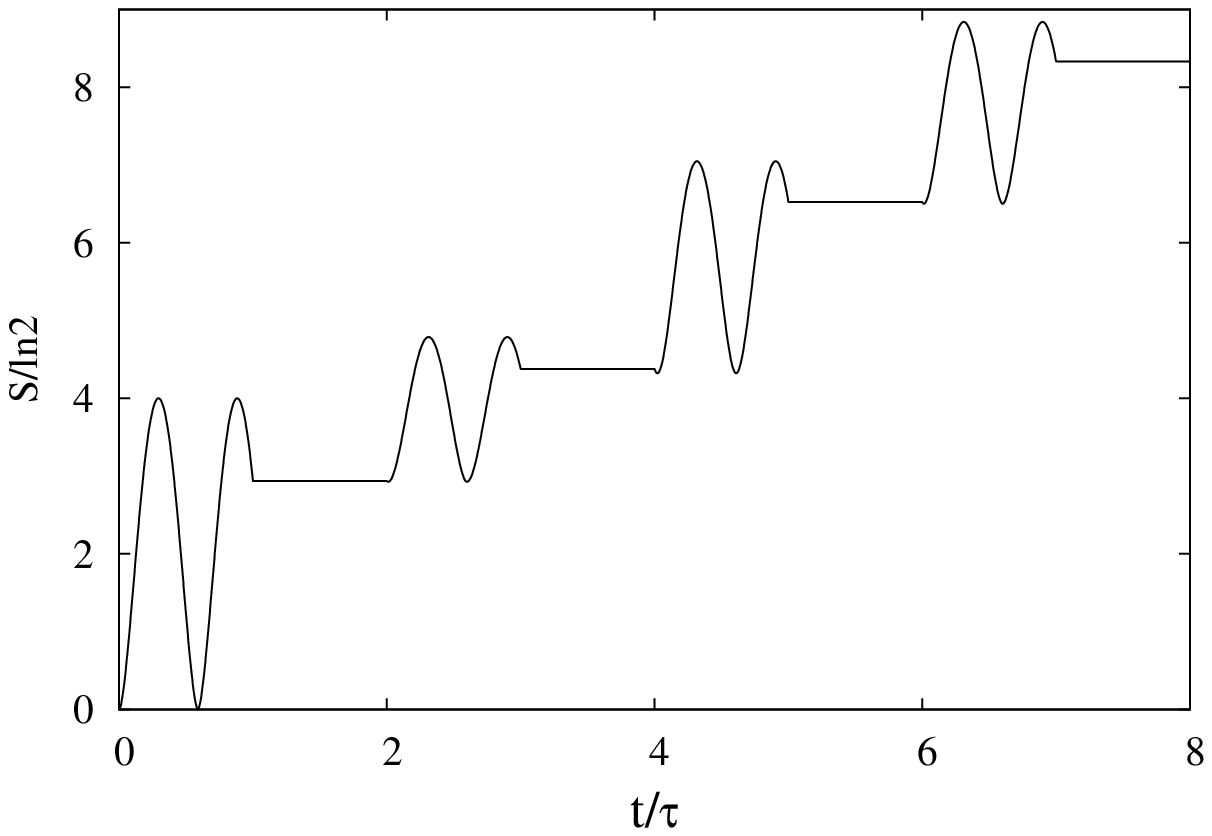}
\hspace{0.5cm}
\includegraphics[scale=.53]{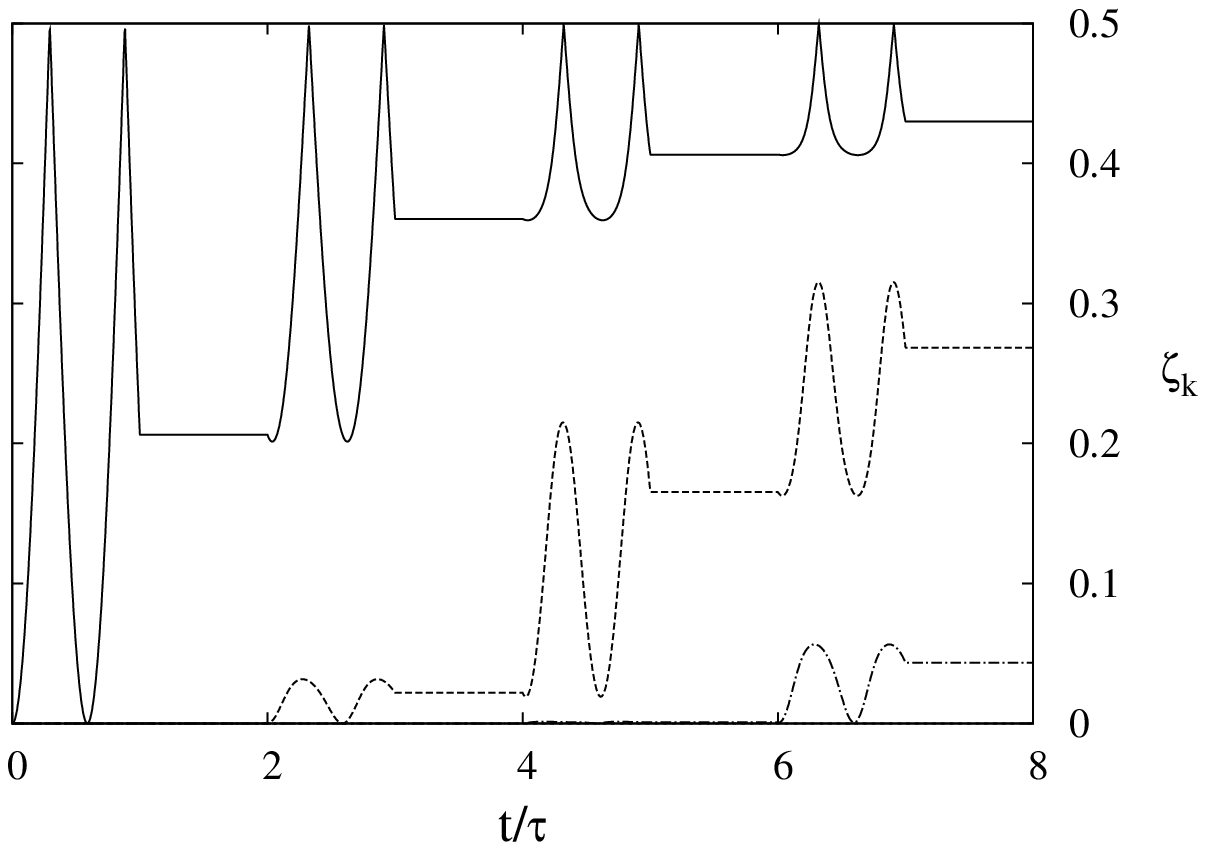}
%\vspace{-0.3cm}
\caption{Left: Entanglement entropy for $S(t)$ for $L=40$ and $\tau= 1.7\pi$.
Right: The eigenvalues $\zeta_k(t)$ producing $S(t)$.}  
\label{fig:zetafulld}
\end{figure}

Let us now discuss the fine structure of the curves. First of all, the increase
takes place in steps with well defined plateaus. These plateaus still occur because 
in one of the two half-periods the time evolution in the subsystem is decoupled from 
that in the surrounding. Thus the entanglement cannot change. 
For small $\tau$, the plateaus are connected by simple monotonous curves, but for larger 
periods, an overshooting occurs which turns into an oscillation for even larger $\tau$.
A particle then performs more than one cycle in a cell before the next switching
occurs. This case is shown in more detail in Fig. \ref{fig:zetafulld} for 
$\tau = 1.7\pi$. The first oscillation is given by the function  
(\ref{eq:deltast}) up to the time $t=\tau$ when the plateau starts. The next 
oscillations look similar but their amplitudes are not the same and even after a
rescaling there remain small differences. In fact, their similarity is even somewhat 
surprising. On the right of Fig. \ref{fig:zetafulld} the non-trivial eigenvalues 
$\zeta_k$ which determine the variation of $S(t)$ are plotted. One sees that 
only one such eigenvalue (and its partner $1-\zeta_k$) exists in the beginning
but that more and more appear in later periods and contribute to $S$. Thus one
expects differences in the functional form.
%
%%%%%%%%%%%%%%%%%%%%%%%%%%%%%%%%%%%%%%%
%%%%%%%%%%%%%%%%%%%%%%%%%%%%%%%%%%%%%%%
%
\par
\begin{figure}[htb]
\center
\includegraphics[scale=.65]{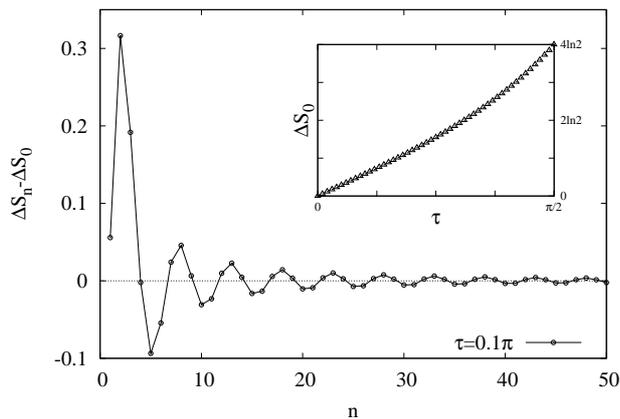}
%\vspace{-0.3cm}
\caption{Variation of the step height in $S(t)$ for $L=200$ and
$\tau= 0.1\pi$. The average value $\Delta S_0$ is subtracted.
Inset: $\Delta S_0$ as a function of $\tau$.}
\label{fig:stepsfulld}
\end{figure}

Looking at the step heights in Fig. \ref{fig:zetafulld}, one sees that they are not
all equal. This is a general feature and illustrated in Fig. \ref{fig:stepsfulld}.
The main graph shows the step height $\Delta S_n$ of the $n$-th step for the initial
increase of $S(t)$ with the average value subtracted. It is an oscillatory decaying 
function independent of $L$ and can be very well described by the expression
\eq{
\Delta S_n = \Delta S_0 + A\frac{\sin(4\tau n - \varphi)}{n^{3/2}}
\label{eq:stepheight}}
with an amplitude $A$ and a phase $\varphi$.
%
%%%%%%%%%%%%%%%%%%%%%%%%%%%%%%%%%%%%%%%
%%%%%%%%%%%%%%%%%%%%%%%%%%%%%%%%%%%%%%%
%
\par
This result is easy to understand if one takes into account that the integrals 
(\ref{eq:fgm}) for the elements of the correlation matrix contain the number
of the period
in the form of factors $\sin(2\gamma_q n)$ and $\cos(2\gamma_q n)$ and that the
main contribution comes from the region near $q=0$ where the dispersion curve has its
maximum. This gives oscillations in $n$ with frequency $2\gamma_0 = 4\tau$ which
persist in $S(t)$. Moreover, the density of states in the integration leads to
a dependence $1/n^{3/2}$, which was also found in a previous treatment of the 
Hamiltonian limit \cite{EP07} and in a calculation of a related TI correlation
function \cite{Prosen00}. The oscillations go away if either  $\tau= 0$ or
$\tau= \pi/2$. Between these two limits, the average step height increases 
monotonously from zero to
$4\ln2$ as shown in the inset of the Figure. Thus the case $\tau= \pi/2$
studied in the last subsection has not only regular steps but also the highest
ones and thus leads to the fastest increase of $S$. Since the dispersion relation 
$\gamma_q$ is invariant under $\tau \rightarrow \pi-\tau$ and under 
$\tau \rightarrow \tau+k\pi$, one can use the above results to obtain the step
heights for arbitrary periods.
\par
In the region beyond the cross-over point at time $L/2v_{\rm max}$ the step
height decays rapidly towards zero with the same $n^{-3/2}$ power law for
large $n$ as before and one can still observe the same oscillations.

\section{Partially dimerized case}
\label{sec:partdim}

\par
\begin{figure}[htb]
\center
\includegraphics[scale=.75]{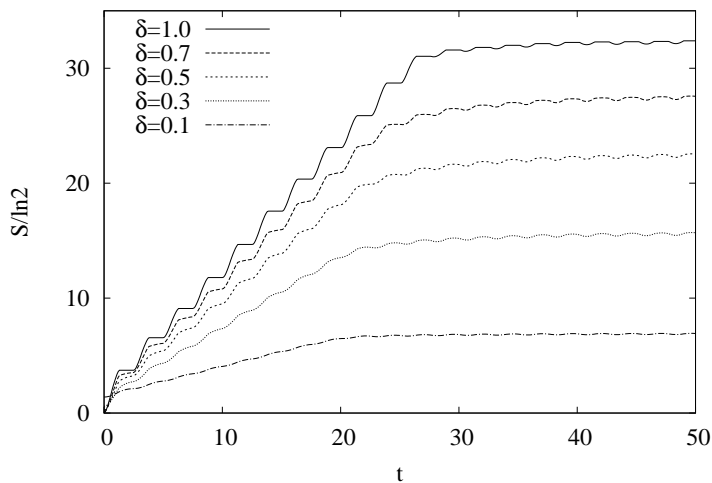}
%\vspace{-0.3cm}
\caption{Entanglement entropy for $L=40$ and $\tau=0.4\pi$ for five
values of the dimerization $\delta$.}
\label{fig:entpartdtausmall}
\end{figure}

We now turn to the case of general values $\delta < 1$, where the system 
has alternating weak and strong hopping. Results for a short quench time, 
$\tau= 0.4\pi$, are shown in Fig. \ref{fig:entpartdtausmall}. The curves
resemble those in Fig. \ref{fig:entfulldtau}, but there is one basic difference.
Since the cells are always coupled, one does not have plateaus any more.
Instead, $S(t)$ increases slowly during those half-periods where the plateaus
occur for $\delta = 1$ and faster for the others. Together this still gives
a kind of step structure, but the steps are washed out.
As $\delta$ becomes smaller, the differences between the two half-periods
also become smaller and the curve smoothens more and more.   
At the same time the 
average slope decreases. Moreover, the initial value $S(0)$ is non-zero and
becomes larger, approaching the value $S = 1/3 \ln L + k$ of the homogeneous
chain. The oscillations with the quench period persist also beyond the bending
point and decrease there with $\delta$ in the same way.

%%%%%%%%%%%%%%%%%%%%%%%%%%%%%%%%%%%%%%%
%%%%%%%%%%%%%%%%%%%%%%%%%%%%%%%%%%%%%%%
%
\par
\begin{figure}[htb]
\center
\includegraphics[scale=.75]{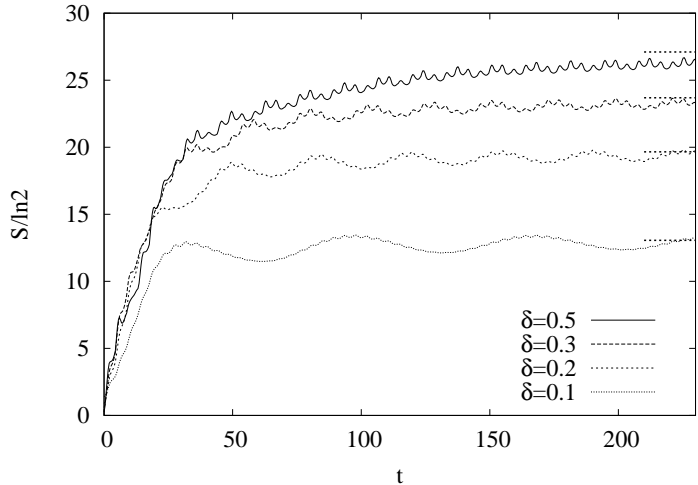}
%\vspace{-0.3cm}
\caption{Entanglement entropy for $L=40$ and $\tau=0.7\pi$ for several
values of the dimerization $\delta$. The dotted lines on the right indicate
the asymptotic values.}
\label{fig:entpartdtaularge}
\end{figure}

This might suggest that the behaviour of $S$ becomes less interesting for 
partial dimerization.
However, this is not true. In Fig. \ref{fig:entpartdtaularge} we show
results for a larger quench time, $\tau= 0.7\pi$ and several values of
$\delta$ below 0.5. In addition to small and fast oscillations one now
observes very slow ones with rather large amplitudes and a frequency which
decreases with $\delta$. As in the previous section, the origin of these
oscillations can be found in the dispersion curve for $\gamma_q$. 
%
%%%%%%%%%%%%%%%%%%%%%%%%%%%%%%%%%%%%%%%
%%%%%%%%%%%%%%%%%%%%%%%%%%%%%%%%%%%%%%%
%
In Fig. \ref{fig:nuq_delta} we have plotted $\gamma_q$ given by Eqs. 
(\ref{eq:gammaq}) and (\ref{eq:cosphi12}) in the Appendix as a function of
$\delta$. One sees that while there is a single maximum at $q=0$ for large
$\delta$, a double-peak structure develops below $\delta = 0.7$ and the
maxima approach the zone boundary at $\pi$ for $\delta \rightarrow 0$.
The reason is that for small $\delta$ the factor $\cos(\varphi_1-\varphi_0)$ 
is almost constant and $\gamma_q$ is basically determined by $\omega_q$ 
in the trigonometric functions. For $\delta = 0$ one finds exactly  
$\gamma_q = 2 \tau\cos(q/2)$ which gives the uppermost curve in the figure
when folded into the first Brillouin zone.
%
%%%%%%%%%%%%%%%%%%%%%%%%%%%%%%%%%%%%%%%
\begin{figure}[htb]
\center
\includegraphics[scale=.6]{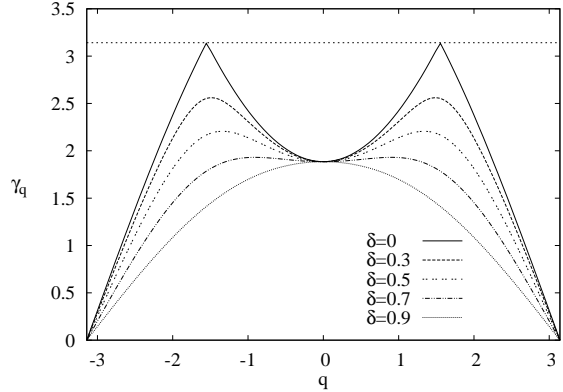}
%\vspace{-0.3cm}
\caption{Dispersion $\gamma_q$ for $\tau=0.7\pi$ and several values of the dimerization.}
\label{fig:nuq_delta}
\end{figure}
%%%%%%%%%%%%%%%%%%%%%%%%%%%%%%%%%%%%%%%
\par
Writing  $\gamma_{\rm max} = \pi-\Delta$, the maxima lead to a long-time
behaviour of the correlations of the form
\eq{
\cos(2n \gamma_{\rm max}) = \cos(2n (\pi-\Delta)) =  \cos(2n \Delta)
\label{eq:gammamax}}
and thus to oscillations with frequency $2 \Delta$.
The effect is the same as in Section \ref{sec:fulldim} for the case
$\tau$ near $\pi/2$ and may be called
an Umklapp effect in the frequency space due to the discrete times. 
The double-peak structure in the dispersion can also lead to beats in $S(t)$, 
if the two frequencies at the maximum and the minimum are close to each other.
Such phenomena have also been observed in \cite{LS05}.

\par
\begin{figure}[thb]
\center
\includegraphics[scale=.7]{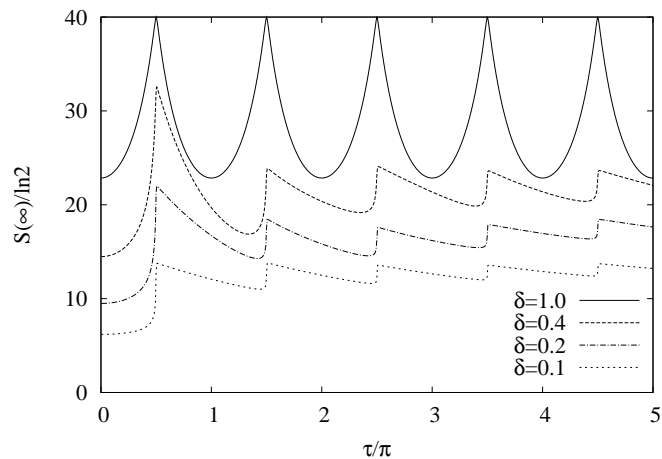}
%\vspace{-0.3cm}
\caption{Asymptotic value of the entanglement entropy for $L=40$ as a function of
the quench period for different dimerizations.}
\label{fig:entasympt}
\end{figure}

Finally, we show in Fig. \ref{fig:entasympt} how the asymptotic value of $S$ depends on 
the parameters $\tau$ and $\delta$. The calculation uses the expressions 
(\ref{eq:fgminf}) in the Appendix for the elements of the correlation matrix
which is then diagonalized. In all cases, $S(\infty)$ is proportional to $L$.
In the fully dimerized case, it is a periodic function of $\tau$ with period 
$\pi$ and maxima at odd multiples of $\pi/2$. The latter times still play a
special role for partial dimerization because the maximal value of $\omega_q$
is always 1, but $S(\infty)$ has only steps there.
In general, it decreases as one lowers $\delta$, which is reasonable, as then
the changes during the quench become smaller. For $\delta=0$, there is no
change at all and $S$ has to be constant in time and non-extensive. One should
note that the asymptotic values give no information on how fast they are reached. 
An extreme case is $\delta=1$ and $\tau$ near $k \pi$. Then the
approach is very slow and if $\tau$ equals $k \pi$ exactly, $S$ remains at its
initial value apart from periodic excursions as noted in Section \ref{sec:fulldim}.
 
%%%%%%%%%%%%%%%%%%%%%%%%%%%%%%%%%%%%%%%%%%%
%%%%%%%%%%%%%%%%%%%%%%%%%%%%%%%%%%%%%%%%%%

\section{Summary and conclusion}

We have studied an integrable quantum chain under the influence of a periodic quench. 
Because one switches back and forth between the two equivalent broken-symmetry 
arrangements, the time evolution is on average that of a translationally
invariant system. The entanglement was studied via the entropy $S$ for a segment
of $L$ sites. This contains more information than the $Q$ measure related to reduced
single-site density matrices and used in \cite{LS05}. We found an overall behaviour
similar to that in a single critical quench, namely an increase of $S(t)$ to a final
value which is proportional to the size of the subsystem. On a finer scale, the
increase is seen to result from the time evolution within the quench periods which
in the simplest case gives a step structure, but becomes partly oscillatory for longer
periods and washed out if the dimerization is weak. On the scale of many periods,
there are in general oscillations related to the spectrum of the evolution operator.
They make the step heights slightly non-uniform and can lead to a striking wave-like 
approach of $S$ to its asymptotic value. A particularly interesting special case
was found for $\delta=1$ and $\tau = \pi/2$, where the electrons move like massless
relativistic particles. This leads to very simple formulae and gives an elementary 
realization of the picture developed by Calabrese and Cardy for the entanglement
process.
\par
We have stressed the close connections to the equilibrium statistics of 
two-dimensional lattice models. These were also visible in the studies of the
kicked Ising chain \cite{Prosen00,LS05,BMM05} but, somewhat surprisingly, not
pointed out explicitly. The equilibrium models provide a useful reference
frame. Thus the concept of the Hamiltonian limit, developed for very anisotropic
equilibrium systems, could directly be taken over. On the other hand,
the unitary time evolution also introduces some differences. The correlations are 
typically oscillating and the discrete time intervals between the periods give rise 
to lattice effects in the frequency domain. The quench could be classified
as critical for all values of the period. Any modification in the relative length
of the half-periods or in the two dimerizations would lead to gaps in the energies
$\nu_q$ and thus to a non-critical quench. The entanglement in this case would
still show similar overall behaviour due to the quasiparticles in $\bar H$ which
propagate with the maximum velocity. However, the various band edges in the
single-particle dispersions will make the picture more complex in detail. Similarly,
a finite quench velocity would complicate matters since in our case one is permanently
crossing the quantum critical point $\delta=0$. The entanglement in such a crossing 
has so far only been studied for single quenches \cite{CherngLev06,Cincio07}. There 
$S$ is not extensive but saturates for large $L$ at a value depending on the transition 
time. It might be interesting to extend the considerations to the periodic case. 
In the framework of equilibrium models, this would correspond to a lattice with slowly 
varying periodic layering.
\par\noindent
\emph{Note added:} A similar situation as studied here was investigated
for Heisenberg chains in \cite{Barmettler08}.

\section*{Acknowledgement}

We would like to thank Thierry Platini and Ulrich Schollw\"ock for discussions
and Peter Barmettler for correspondence.

\section*{Appendix: Calculation of the correlation matrix}

In this Appendix we summarize the method of obtaining the 
matrix $\mathbf{C}(t)$ from which the entanglement
entropy is calculated for arbitrary dimerizations and quench periods.
\par
In order to calculate the time evolution of the fermionic
operators $a_j$ and $b_j$, it is convenient to switch to the
Fourier-transformed operators defined in (\ref{eq:ab_ft}).
Then the $a_q$ and $b_q$ evolve for each $q$ independently as
\eq{
\left(\begin{array}{c} a_q(t) \\ b_q(t) \end{array}\right) =
\left(\begin{array}{cc}
\cos{\omega_q t} & i \ee^{-i\varphi_q}\sin{\omega_q t}\\ 
i \ee^{i\varphi_q}\sin{\omega_q t} & \cos{\omega_q t}
\end{array}\right)
\left(\begin{array}{c} a_q \\ b_q \end{array}\right) =
v_q(t)\left(\begin{array}{c} a_q \\ b_q \end{array}\right)
\label{eq:abt}}
which is the analogue of (\ref{eq:abtfull}). In Fourier space, 
the full evolution matrix always decomposes into such blocks.
\par
During the quench the system is evolved with two different Hamiltonians
$H_0$, $H_1$ with corresponding matrices $v_{0q}(\tau)$, $v_{1q}(\tau)$
and the time dependent operators after $n$ full periods are given through
\eq{
u_q(2n\tau)=\left[ v_{0q}(\tau)v_{1q}(\tau) \right]^n.
\label{eq:uq}}
In order to obtain $u_q(2n\tau)$ one has to raise the matrix $u_q(2\tau)$
to the $n$-th power, thus one has to determine its eigenvalues and
eigenvectors, as well. The eigenvalues can again be written as
$\exp(\pm i\gamma_q)$ where
\eq{
\cos \gamma_q = \cos^2 \omega_q \tau - \cos(\varphi_1-\varphi_0)
\sin^2 \omega_q \tau
\label{eq:gammaq}}
with
\eq{
\cos(\varphi_1-\varphi_0) =  \frac{\cos^2(q/2)-\delta^2 \sin^2(q/2)}
{\cos^2(q/2)+\delta^2 \sin^2(q/2)}\,.
\label{eq:cosphi12}}
To simplify the notation we have dropped the $q$ indices of $\varphi_0$
and $\varphi_1$. Denoting the first and second component of the $i$-th
eigenvector by $A_i$ and $B_i$, respectively,
the equations then yield
\eq{
|A_{1,2}|^2=\frac 1 2 \left[ 1 \mp 
\frac{\sin(\varphi_1-\varphi_0)\sin^2 \omega_q\tau }
{\sin \gamma_q} \right]
\label{eq:evec1}}
and
\eq{
B_{1,2} = \frac{\sin \omega_q\tau \cos \omega_q\tau \,
(\ee^{i\varphi_1}+\ee^{i\varphi_0})}
{\pm\sin \gamma_q - \sin(\varphi_1-\varphi_0)
\sin^2 \omega_q\tau} \, A_{1,2} \; .
\label{eq:evec2}}
With the help of (\ref{eq:gammaq})--(\ref{eq:evec2}) one can construct
the matrix $u_q(2n\tau)=u_q^n(2\tau)$ whose elements read
\eq{
\begin{split}
u_{11} & = \cos n\gamma_q - i \frac{\sin n\gamma_q}{\sin \gamma_q}
\sin(\varphi_1-\varphi_0) \sin^2 \omega_q\tau \\
u_{12} & = i \frac{\sin n\gamma_q}{\sin \gamma_q}\,
(\ee^{-i\varphi_1}+\ee^{-i\varphi_0}) \sin \omega_q\tau \cos \omega_q\tau
\end{split}
\label{eq:uij}}
while the other two elements $u_{21}=-u^*_{12}$ and $u_{22}=u^*_{11}$ 
follow from unitarity.
\par
Finally, one has to calculate the correlation matrix elements
$C_{lj}(2n\tau)$ as defined in (\ref{eq:correlt}) which requires
knowledge of the initial correlations. These are given as
\eq{
\langle a^\dag_q a_q\rangle = \langle b^\dag_q b_q\rangle = 1/2 \; ,
\quad
\langle a^\dag_q b_q\rangle = \langle b^\dag_q a_q\rangle^* 
= \ee^{i\varphi_0} / 2 \; .
\label{eq:abq0}}
Going over to Fourier transforms and using the
form of the evolution matrix $u_q(2n\tau)$ one obtains
$C_{lj}(2n\tau)=\frac 1 2 (\delta_{lj}+\Gamma_{lj})$ with the block matrix
\eq{
\mathbf{\Gamma}=
\left(\begin{array}{cccc}
\Pi_0 & \Pi_{-1} & \cdots & \Pi_{1-N} \\
\Pi_1 & \Pi_0 & & \vdots \\
\vdots & & \ddots & \vdots \\
\Pi_{N-1} & \cdots & \cdots & \Pi_0
\end{array}\right) \; ,
\quad\quad
\Pi_m=
\left(\begin{array}{cc}
f_m & g_m \\
g^*_{-m} & -f_m
\end{array}\right)
}
where $m=l-j$ and the matrix elements of the $2 \times 2$ matrices 
$\Pi_m$ read in the thermodynamic limit
\eq{
\begin{split}
f_m & =  \int_{-\pi}^{\pi} \frac{\dd q}{\pi}
\mathrm{Re}\left( u_{22}u_{12}\ee^{i\varphi_0}\right)\ee^{-iqm} \, , \\
g_m & = \int_{-\pi}^{\pi} \frac{\dd q}{2\pi}
\left( u^2_{22}\ee^{i\varphi_0} - u^2_{21}\ee^{-i\varphi_0} \right)\ee^{-iqm}
\, .
\end{split}
\label{eq:fgm}}
\par
Up to now we have considered only the special time instances
$t_n=2n\tau$. To extend the calculation to arbitrary
times $t_n<t<t_{n+1}$ one has to calculate the matrix elements of
\eq{
u_q(t)=
\begin{cases}
v_{1q}(t-t_n)\, u_q(t_n), & \text{if } t<t_n+\tau \\
v^+_{0q}(t_{n+1}-t)\, u_q(t_{n+1}), & \text{if } t \ge t_n+\tau
\end{cases}
\label{eq:uqt}}
and substitute them in (\ref{eq:fgm}) to obtain the matrix
$\mathbf{\Gamma}$ for intermediate times.
\par
It is also interesting to investigate the asymptotic ($n\to\infty$)
behaviour of the matrix elements. Then the integrals in (\ref{eq:fgm}) have
rapidly oscillating arguments, and one can replace
$\cos n\gamma_q$ and  $\sin n\gamma_q$ by zero as well as
$\cos^2 n\gamma_q$ and $\sin^2 n\gamma_q$ by $1/2$, as these factors
are multiplied by smooth functions of $q$ and integrated over.
After a lengthy calculation one has the form
\eq{
\begin{split}
f_m(\infty) & =  \int_{-\pi}^{\pi} \frac{\dd q}{2\pi}
\frac{i \sin \omega_q\tau \cos \omega_q\tau}
{\delta^2 \tan^2 \frac q 2 + \cos^2 \omega_q\tau}
\delta \tan \frac q 2 \sin qm  \, , \\
g_m(\infty) & = \int_{-\pi}^{\pi} \frac{\dd q}{2\pi}
\frac{\omega_q\cos^2 \omega_q\tau}
{\delta^2 \tan^2 \frac q 2 + \cos^2 \omega_q\tau}
(\cos qm + \tan \frac q 2 \sin qm) \, .
\end{split}
\label{eq:fgminf}}
For $\delta = 1$ and $\tau \rightarrow 0$ all expressions simplify
considerably. Then $f_m$ and $g_m$ become Bessel functions and their
asymptotic values are $f_m(\infty)=0$ and $g_m(\infty)=(\delta_{m,0}+\delta_{m,1})/2$.
This gives the correlation matrix found in \cite{EP07} for a single
quench.
%

%\bibliographystyle{adp}

%\bibliography{perquench_refs}

\providecommand{\WileyBibTextsc}{}
\let\textsc\WileyBibTextsc
\providecommand{\othercit}{}
\providecommand{\jr}[1]{#1}
\providecommand{\etal}{~et~al.}

\end{document}